\documentclass[aps,twocolumn]{revtex4} 
\usepackage{amssymb,amsmath,amscd}
\usepackage{graphicx} 

\def\a={&=&}
\def\beqa{\begin{eqnarray}}
\def\eeqa{\end{eqnarray}}
\def\nnn{\nonumber \\} 
\def\ne{\nu_{\!e\!f\!f}}
\def\oe{\omega_{e\!f\!f}}
\def\cO{{\cal O}}
\def\vr{{\vec{r}}}
\def\exy{{\eta(x,y)}}
\def\exyx{{\eta\big(x,y(x)\big)}}
\def\exyt{{\tilde{\eta}_{\yL,\yU}}}
\def\pglu{{P^0_{\yL,\yU}}}
\def\Wa{W}
\def\Wao{W_0}
\def\xm{x\!\!-\!\!1}
\def\xp{x\!\!+\!\!1}
\def\ym{y\!\!-\!\!1}
\def\yp{y\!\!+\!\!1}
\def\yma{y_{\max}}
\def\ymi{y_{\min}}
\def\yL{y_{_L}}
\def\yU{y_{_U}}

\newcommand{\pd}[2]{\frac{\partial {#1}}{\partial {#2}}}
\newcommand{\av}[1]{\overline{#1}}
\newcommand{\ave}[1]{\langle {#1} \rangle}
\newcommand{\AVE}[1]{\left\langle {#1} \right\rangle}
\def\TwoFigPlace#1#2#3#4#5#6#7#8#9
{{\hbox{} \vbox to #3 true in {}
  \includegraphics{#1}
  \includegraphics{#2}
}}

\begin{document}
\title[]{
Directed polymers in random media under confining force
}
\author{Hyeong-Chai Jeong}
\email{hcj@sejong.ac.kr} 
\affiliation{
Department of Physics
and Institute for Fundamental Physics, Sejong University, 
Seoul 143-747 \\
and Asia Pacific Center for Theoretical Physics,
POSTECH, Pohang 790-784, Korea
}
\received{\today}
 
\begin{abstract}
The scaling behavior of a directed polymer in a two-dimensional (2D)
random potential under confining force is investigated.
The energy of a polymer with configuration $\{y(x)\}$
is given by $H\big(\{y(x)\}\big) = \sum_{x=1}^N \exyx + \epsilon
\Wa^\alpha$, where $\eta(x,y)$ is an uncorrelated random potential and
$\Wa$ is the width of the polymer. Using an energy argument, it is
conjectured that the radius of gyration $R_g(N)$ and the energy
fluctuation $\Delta E(N)$ of the polymer of length $N$ in the ground
state increase as $R_g(N)\sim N^{\nu}$ and $\Delta E(N)\sim N^\omega$
respectively with $\nu = 1/(1+\alpha)$ and $\omega =
(1+2\alpha)/(4+4\alpha)$ for $\alpha\ge 1/2$. A novel algorithm of
finding the exact ground state, with the effective time complexity of
$\cO(N^3)$, is introduced and used to confirm the conjecture
numerically.  
\end{abstract}

\pacs{PACS Numbers: 72.15.Rn} 

\maketitle

\section{Introduction}
Scaling behaviors of a
directed polymers in a random media (DPRM) have been studied extensively
due to the mathematical interest of optimal path problems in random
environment in addition to their application to physical systems 
such as stretched polymer in a gel or flux lines in a disordered
superconductor~\cite{Barabasi95B}. 
A directed polymer is stretched in the longitudinal
direction but can fluctuate in the transversal direction.
We can map a configuration of a directed polymer in a $d$-dimensional
space to a path of random walkers in the $d_t = d-1$ dimensional
space (in the transversal direction) when the longitudinal direction 
is considered as the time. 
Therefore, in sufficiently low temperature, a DPRM corresponds to the
minimum energy path of a random walker in a random potential
$\eta(t,\vr)$. Its scaling properties are characterized by two exponents, 
transverse length scale exponent $\nu_o$ and energy fluctuation 
exponent $\omega_o$~\cite{Huse85PRL}.
For the $d_t=1$ dimensional walkers, the exact values of these 
exponents, $\nu_o=2/3$ and $\omega_o=1/3$ can be obtained analytically 
by the combinatorial method~\cite{Johansson00}.
The exact exponent values can also be obtained by 
mapping the free energy of a random walker to 
Kardar-Parisi-Zhang equation (KPZ) via Cole-Hope
transformation~\cite{Kardar87,Kim9108,Krug92,Halpin95,Barabasi95B}.
This mapping shows that the exponents for the DPRM, $\nu_o$ and
$\omega_o$ should be equal to the inverse of the dynamic exponent, 
$1/z_{_{KPZ}} = 2/3$ and the growth exponent, $\beta_{_{KPZ}} = 1/3$
of KPZ equation respectively at any finite temperature. 
In other words, any disorder drives a directed polymer in the 
$d=2$ dimensional media into 
a strong disorder, pinned phase at $T=0$.

\begin{figure}[t!] 
\includegraphics[width=8cm]{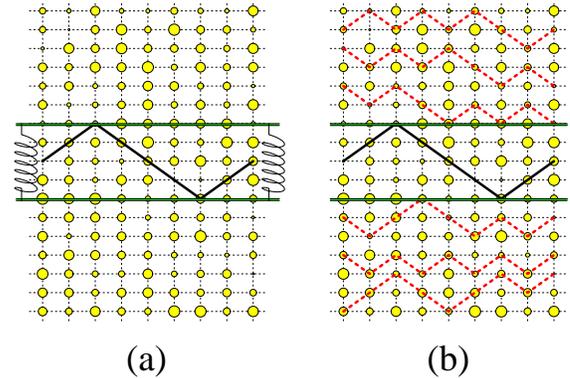} 
\caption[0]{(Color online) Directed polymers in 2D random media 
under confining force. The magnitude of the random potential 
at each lattice site is denoted by the area of the circles 
at the lattice site. A directed polymer is confined by
the two rods. The confining force may come from 
the elasticity which are symbolized by the spring~(a)
or from the repulsion of the polymers (red dashed lines) at
the outside of the rods~(b). 
}
\label{fig.dprm.cf}
\end{figure}

In this paper, we study the scaling properties of a DPRM under 
confining force (DPRMCF). The Hamiltonian for a DPRMCF has two
terms, the usual random potential term $E_{RM}$
of the DPRM and the confining energy term $E_C$
which prefers the straight polymer in global length scale.
It may describe a DPRM confined by an
inflatable but non-flexible tube or by two rods with springs as
illustrated in Fig.~\ref{fig.dprm.cf}(a).  
We consider the confining energy which depends only on the ``global
width'' $W$ of the polymer. Such confining energy term may mimic
the elastic energy of the (inflatable) tube or the spring 
which prefers the smaller width.
The confining force on the center polymer may arise 
from the repulsion from the other polymers (red dashed line)
outside of the rods if we consider an array of polymers 
on a plate as illustrated in Fig.~\ref{fig.dprm.cf}(b). 
We may find other physical systems that our model 
might be applied, such as the motion of a single step on 
a terrace with quenched random impurities in a vicinal
surface~\cite{Jeong99SSR} but the principal motivation for the model
is rather theoretical interest.
The effects on the scaling properties of random walkers or directed
polymers from the energy terms associated with the global
configuration quantity, such as the global width $\Wa$, are
theoretically intriguing~\cite{Noh01PRE,Kim02,Jeong02PRE,Jeong03PRE}. 
For example, the confining energy term $E_c \sim W$ makes the normal
random walkers (without $E_{RM}$) visit the same $y$ value sites even
times stochastically~\cite{Noh01PRE}, and changes the ``roughness''
exponent to 1/3 from the conventional universal value of
$1/2$~\cite{Noh01PRE,Jeong02PRE,Kim04JKPS}. 
The path of this stochastic even visiting random walker
can be mapped a 1D interface profile which is called a
``self-flattening'' surface~\cite{Kim02,Jeong03PRE}. 
The height-height correlation function of the self-flattening surface
shows an anomalous scaling behavior in the sense that 
the local wandering exponent is different
from the global roughness exponent~\cite{Jeong03PRE}.
This anomalous behavior indicates the existence of window
length scale $l(N) \sim N^\delta$ with $0< \delta < 1$
above which the global energy term becomes relevant. 
Recently, it has been conjectured that
the window exponent $\delta$ is given by 
$\delta=\frac{D}{D+\nu_o}$ from an energy-entropy
argument where $D$ is the dimension of the surface 
and $\nu_o$ is the roughness exponent without
the self-flattening mechanism~\cite{Park03}. 
This argument seems to be valid for general self-flattening
mechanism, which corresponds to non-zero 
finite temperature dynamics with the confining energy 
proportional to the width $\Wa$~\cite{Kim02,Jeong02PRE,Jeong03PRE}. 

Here, we study the scaling properties of the 
zero temperature, ground state paths for
a DPRMCF in 2D with a confining energy in the form of
$E_c = \epsilon W^\alpha$~\cite{loc-cw}. 
If we assume the existence
of window length $l(N)\sim N^\delta$, a power counting 
argument predict $\delta = 3/(2+2\alpha)$ for $\alpha>1/2$ 
and the confining energy of any positive $\epsilon$ changes
the roughness exponent $\nu$ to $1/(1+\alpha)$ from 
$\nu_o = 2/3$ and the energy fluctuation exponent $\omega$ to 
$(1+2\alpha)/(4+4\alpha)$ from $\omega_o=1/3$. This conjecture is
numerically tested for a series of different $\alpha$ values.

This paper is organized as follows. In Sec.~II, we 
define the DPRMCF model. An analytic prediction for the roughness
exponent $\nu$ and the energy fluctuation exponent $\omega$ is
presented in Sec.~III. A novel algorithm for a numerical study of
DPRMCF which finds the exact ground state in the effective time
complexity of $\cO(N^3)$ is introduced in Sec.~IV. In Sec.~V, we 
present the numerical results from a series of Monte Carlo Simulations
for DPRMCF model and calculate the exponents $\nu$ and $\omega$. We 
conclude the paper with a summary and remarks in Sec.~VI.

\section{Model}
We consider a discrete model for a directed polymer in a random media
under confining force (DPRMCF) whose Hamiltonian has two terms, site 
random potential 
and confining energy.  Before introducing the details of our DPRMCF model,
let us first explain a discrete model for DPRM without the confining 
force. We consider polymers on a 
2D lattice  whose horizontal axis is $x$ and 
the vertical axis is $y$ ({\it i.e.}, the $d_t=1$ dimensional
transversal space is the $y$-axis).
At each lattice site, random potential $\eta(x,y)\in[0,1]$
taken from a uniform distribution with 
\beqa
 \AVE{\eta(x,y)\,\eta(x',y')}
	\a= \frac{1}{4}+\frac{1}{12} \delta_{x,x'}\,\delta_{y,y'}
\eeqa
is assigned. 
Since overhangs are forbidden for a {\it directed} polymer,
its configuration is specified by 
a single valued integer function $y(x)\in{\mathbb Z}$ 
with $|y(\xp)-y(x)|$=1.
The energy of a DPRM is given by the
sum of site potentials $\exyx$ on which the polymer lies.
At zero temperature, a polymer on a given random potential 
locates at the path which minimizes the total site energies of the
random media,
\beqa
 E_{RM}\big(\{y(x)\}\big) \a= \sum_{x=1}^N \exyx,
\label{eq.erm}
\eeqa
where $N$ is the length of the polymer. 

The configurational space of a DPRMCF is the same as that 
of a DPRM but the Hamiltonian $H$ of a DPRMCF
has the confining energy term $E_{C}=\epsilon\, \Wa^\alpha$
in addition to the $E_{RM}$,
\beqa
 H(\{y(x)\}) = \sum_{x=1}^N \exyx + \epsilon\, \Wa^\alpha.
\label{eq.h}
\eeqa 
Here, $\Wa$ is the absolute ``width'' of the polymer defined by
\beqa
 \Wa	\a= y_{\max} - y_{\min} + 1,
\label{eq.W}
\eeqa
where $y_{\max}$ and $y_{\min}$ are the maximum and the minimum values
of $y(x)$ respectively 
and $\epsilon$ and $\alpha$ are positive parameters so that
the confining energy increases when the width grows.

\section{Window hypothesis and power counting argument 
from energy fluctuation}
We study the scaling properties of the 
zero temperature, ground state paths for
the Hamiltonian of Eq.~(\ref{eq.h}). 
The statistical behaviors of a polymer in random 
media are usually characterized by the radius of gyration 
$R_g$ and the energy fluctuation $\Delta E$ defined
by 
\beqa
 &\mbox{} \hspace{-2mm} R_g(N) 
 \ = \sqrt{\AVE{\,\av{(y-\av{y})^2}\,}}	\hspace{-15mm}
 \nnn
 & \mbox{} \hspace{-20mm} = &
 \mbox{} \hspace{-20mm} 
 \sqrt{\frac{1}{\Omega} \sum_{\{\eta\}} 
	\left[ \frac{1}{N} \sum_{x=1}^N 
		\left(y_\eta(x) - \frac{1}{N} \sum_{x=1}^N y_\eta(x)
		\right)^2 
	\right]}
\label{eq.rg}
\eeqa
and
\beqa
 & \mbox{} \hspace{-3mm} \Delta E(N) 
	\ = \sqrt{\AVE{\Big(E - \ave{E} \Big)^2}}  
	\mbox{} \hspace{-30mm}  
\nnn
& \mbox{} \hspace{-10mm} =&
 \mbox{} \hspace{-10mm} 
\sqrt{\frac{1}{\Omega} \sum_{\{\eta\}} 
	\left[ H_\eta(\{y_\eta(x)\}) 
		 - \frac{1}{\Omega} \sum_{\{\eta\}} 
		H_\eta(\{y_\eta(x)\}) \right]^2}, \nnn
	&
\label{eq.de}
\eeqa
where $\Omega$ is the number of different realizations of
random potentials and $\{y_\eta(x)\}$ and $H_\eta(\{y_\eta(x)\})$
are the ground state path and its energy for 
a given random potential $\eta$. 
These two quantities asymptotically increase as 
\beqa
 R_g(N) \sim N^\nu	\\
 \Delta E(N) \sim N^\omega
\eeqa
for the ground state paths for a variety of different distributions of
the randomness~\cite{Huse85PRL,Hartmann01,Hansen04}.

In this section, we estimate $\nu$ and $\omega$ for the DPRMCF by
comparing the $E_{RM}$ and $E_C$ in the ground state. 
For a directed polymer in a 2D random media, it has been well known
that the energy of the minimum path of length $N$ 
can be asymptotically written as 
\beqa
 E_{DPRM}(N) = a N + b N^{\omega_o}, 
\label{eq.ene.dprm}
\eeqa
with $N$ independent positive parameters,
$a$, $b$ and $\omega_o$.
As $N$ goes to infinity,
the parameter $a$ becomes sample independent
and is about 0.25 for a random potential
$\eta(x,y)\in[0,1]$
taken from a uniform distribution.
On the other hand, the second coefficient $b$
is a sample dependent parameter with a finite
variation. The energy fluctuation of
the DPRM with length $N$ is given by
\beqa
 \Delta E_{DPRM}(N)	
	\a= \sqrt{\AVE{\Big(E_{DPRM} - \ave{E_{DPRM}}\Big)^2}} \nnn
	\a= \sqrt{(\Delta b)^2}\, N^{\omega_o},
\eeqa
and $\omega_o$ is called the energy fluctuation 
exponent for the DPRM~\cite{Kim9108,Krug92}. 

We estimate $\nu$ and $\omega$ for our DPRMCF model using
Eq.~(\ref{eq.ene.dprm}) with an assumption of the existence of
``window''~\cite{Noh01PRE,Jeong03PRE}. The confining energy term 
is assumed to be relevant only beyond a certain length scale
of $l \sim N^\delta$, so called ``window''~\cite{Noh01PRE,Jeong03PRE}
so that the polymer behaves like DPRM within the length scale of $l$. 
Then, the random potential energy $E_{RM}(l;N)$ of DPRMCF with length
$N$ can be written as 
\beqa
 E_{RM}(l;N)
	\a= \left( a l + b l^{\omega_o} \right) (N/l)	\nnn
	\a= a N + b N l^{\omega_o - 1}
\eeqa
and the total energy of DPRMCF is given by 
\beqa
 H(l;N)	\a= E_{RM}(N) + \epsilon \Wa^\alpha	\nnn
	\a= a N + b N l^{\omega_o - 1} 
		+ c l^{\alpha\nu_o},
\label{eq.hln}
\eeqa
where $a$, $b$ and $c$ are independent of $l$ and $N$ 
since the absolute width of the polymer $W$ is of order $l^{\nu_o}$.
From the minimization of $H(l;N)$ with respect to $l$ 
({\it i.e.}, from $\pd{H}{l} =0$),
we have
\beqa
 \alpha \nu_o \delta - \delta	\a=  1 + \omega_o \delta - 2 \delta
\eeqa
and the window exponent $\delta$  
and the roughness exponent $\nu$ are given by 
\beqa
 \delta \a= 1/(1+\alpha\nu_o-\omega_o)	\nnn
	\a= 3/(2+2\alpha),		\\
 \nu	\a= \delta \nu_o	\nnn
	\a= \nu_o / (1+\alpha\nu_o-\omega_o)	\nnn
	\a= 1/(1+\alpha),
\label{eq.nu}
\eeqa
where we use $\nu_o = 2/3$ and $\omega_o = 1/3$ 
for 1D DPRM~\cite{Huse85PRL,Kim9108,Krug92}. 
The above two equation for $\delta$ and $\nu$ are only valid
for $\alpha\ge 1/2$ since $\delta$ cannot be larger than 1. For
$0\le \alpha < 1/2$, we expect $\nu=\nu_0$ since the confining energy
term is not relevant since the third term of Eq.~(\ref{eq.hln}) 
can be always neglected comparing to the second term as $N$ goes 
to infinity.

We can use a similar argument to estimate the energy
fluctuation exponent. We first assume 
that the total energy fluctuation $\Delta H$ should show
the same $N$ dependence as the random potential energy
fluctuation $\Delta E_{RM}$ for $\alpha>1/2$
from the power counting argument. 
Then, for the estimation of $\Delta E_{RM}$,
we use the same ``window'' argument used to extract
the roughness exponent $\nu$.
Since there are $N/l \sim N^{1-\delta}$ windows whose energies
fluctuate with the 
``amplitude'' $l^{\omega_o} \sim N^{\omega_o \delta}$,
we have
\beqa
 \Delta E_{RM}(N)
	\a= N^{\omega_o\,\delta}\  N^{\frac{1-\delta}{2}}
\eeqa
and the energy fluctuation exponent $\omega$ is given by
\beqa
 \omega \a= \omega_o\, \delta + \frac{1}{2} - \frac{1}{\delta}	\nnn
	\a= \frac{1+2\alpha}{4(1+\alpha)}.
\label{eq.omega}
\eeqa

\section{Exact enumeration Algorithm}

\begin{figure}[t!] 
\includegraphics[width=8cm]{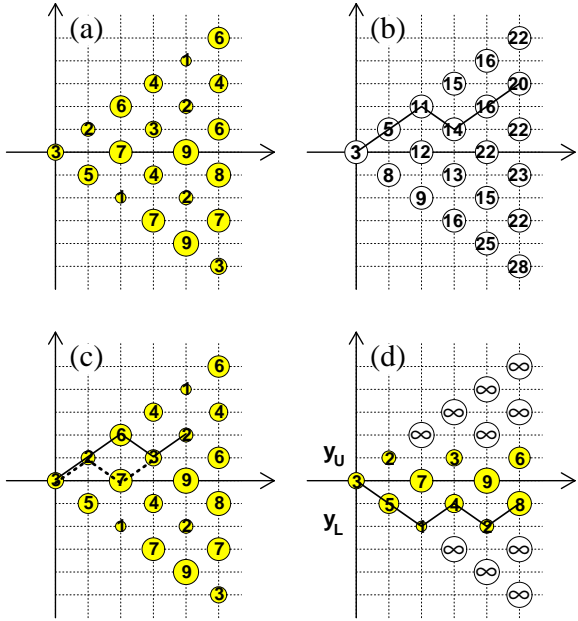} 
\caption[0]{Directed polymer in 2D random media.
(a) A random potential is assigned at each lattice site.
(b) Transfer matrix algorithm for a DPRM. The number
in the circle at the site $(x,y)$ is the sum of 
site potentials of the minimum energy path 
from the origin to the site $(x,y)$.
The solid line is the minimum energy path to $x=N$. 
(c) For the DPRMCF, the minimum energy path to the site
$(x,y)$ is not a simple sum of the site $(x,y)$ and the
minimum energy path to $(\xm,\ym)$ or $(\xm,\yp)$ (see text). 
(d) The minimum energy path for a modified potential 
$\exyt(x,y)$ with $\yL=-2$ and $\yU=1$. 
}
\label{fig.tran.mat}
\end{figure}

In this section, we consider a numerical method to
find the ground state of the DPRMCF. For a given random energy
$\eta(x,y)\in\,[0,1]$ for the 2D lattice sites 
$(x,y)\in {\mathbb Z}^2$, 
we find the minimum energy path of Eq.~(\ref{eq.h}) using an exact
enumeration method under the constraint of $|y(\xp)-y(x)|=1$ with the
anchored \big($y(0)=0$\big) boundary condition. Since the number of
all possible paths increases as $2^N$ for a polymer of length $N$, it
is (computationally) impossible for large $N$ (say, $N>50$) to find
the ground state by comparing the energies of all paths.
We present a novel, efficient algorithm of finding the exact ground
energy path with the effective time complexity $\cO(N^3)$, based on
the transfer 
matrix algorithm (TMA)~\cite{Huse85PRL} for the DPRM problem.
Figure~\ref{fig.tran.mat}(a) and (b) illustrate the TMA for 
the DPRM problem. 
We represent the random potential in (a)
by writing its value at each lattice site 
in the circle at the site
(For clarity, we show an integer-valued random potential 
$\eta(x,y)\in\{1,2,\ldots,9\}$
instead of real numbers from $[0,1]$.) 
and show the minimum energy path in (b) by
a solid line. 
The number in the circle at the site ($x,y$)
in (b) is the minimum potential energy $E^0_m(x,y)$ of a polymer 
from the origin to the site $(x,y)$. 
The minimum energies $E^0_m(x,y)$ for all the $(x,y)$
with $|y|<x$ for $x\in\,\{1,2,\cdots,N\}$ can be obtained
in $\cO(N^2)$ time using the TMA,
\beqa
 E^0_m(x,y) \a= \eta(x,y) 		\nnn
	& + &\min [E^0_m(\xm,\ym), E^0_m(\xm,\yp)]
	\hspace{10mm}
\label{eq.tr}
\eeqa
with $E^0_m(0,0)=\eta(0,0)$ and $E^0_m(x,y)=\infty$ for $|y|>x$, 
where $\min [A,B]$ is the minimum value of $A$ and $B$.
The minimum energy path $P^0_m(x,y)$ from the origin to 
the site $(x,y)$ can be obtained similarly;
\beqa
 P^0_m(x,y) 
	&\hspace{-2mm} = \hspace{-2mm}&
	\left\{\begin{array}{l}
	(x,y) + P^0_m(\xm,\ym) \\
 	\hspace{6mm} \mbox{if\ }  E^0_m(\xm,\ym)< E^0_m(\xm,\yp)\\
	(x,y) + P^0_m(\xm,\yp) \\
 	\hspace{6mm} \mbox{if\ }  E^0_m(\xm,\ym) > E^0_m(\xm,\yp)
	\end{array}
	\right. 
	\hspace{5mm}
\eeqa
with $P^0_m(0,0)=(0,0)$. (If $E^0_m(\xm,\ym) = E^0_m(\xm,\yp)$,
the minimum energy path is not unique and 
both paths become the minimum energy paths, 
but this is a very rare event for real values 
of $\eta$.)
The ground state energy $E^0_g(N)$ is then given by the minimum of 
$E^0_m(N,y)$ over $y$,
\beqa
 E^0_g(N) \a= \min_{y} E^0_m(N,y)
\eeqa
for $|y|\le N$
and the ground state path is given by 
\beqa
 P^0_g(N) \a= P^0_m\left(N,y_m^0(N)\right)
\eeqa
where 
$y^0_m(N)$ is the site such that $E^0_m\big(N,y^0_m(N)\big)=E_g(N)$.
The ground state path for the 
random potential of Fig.~\ref{fig.tran.mat}(a)
is obtained this way and shown in~(b)
as the solid line.

One might suppose that a similar transfer matrix algorithm,
with the time complexity of $\cO(N^2)$,
might be possible for a DPRMCF.
The first term of Eq.~(\ref{eq.tr}), $\eta(x,y)$ is the energy
cost for a DPRM to proceed one step further from $\xm$ to $x$. 
A naive generalization of Eq.~(\ref{eq.tr}) is replacing
$\eta(x,y)$ by 
$\eta(x,y) + \epsilon\left[(\Wa+1)^\alpha - \Wa^\alpha \right]$
when the one-step movement increases the polymer width
from $\Wa$ to $\Wa+1$. 
However, such algorithm does not lead to the global ground 
state of Eq.~(\ref{eq.h}) since the minimum energies 
$E_m(\xm,\ym)$ and $E_m(\xm,\yp)$ are not enough to
determine $E_m(x,y)$ as illustrated in 
Fig.~\ref{fig.tran.mat}(c) where $E_m(x,y)$ is
the energy (including the confining term) of the
ground state path to the point $(x,y)$ from the origin. 
As an example, consider the minimum energy path
to the point (4,2) $P_m(4,2)$
for $\alpha=1$ and $\epsilon=2$
with $\eta(x,y)$ given in Fig.~\ref{fig.tran.mat}(a).
By comparing energies of all paths, one can see that 
$P_m(4,2)$ is given by
$P_m(4,2)$=(0,0)-(1,1)-(2,2)-(3,1)-(4,2), denoted by
the solid line in Fig.~\ref{fig.tran.mat}(c).
This path is not a simple addition of
the end point (4,2) to $P_m(3,1)$ unlike
the case of the DPRM. The minimum energy path to (3,1)
is given by $P_m(3,1)$=(0,0)-(1,1)-(2,0)-(3,1), 
denoted by the dashed line. It has lower energy 
than $P_e(3,1)$ = (0,0)-(1,1)-(2,2)-(3,1) 
although $\eta(2,0)>\eta(2,2)$ since
the global width of $P_m(3,1)$ is smaller than
that of $P_e(3,1)$. 
However, $P_m(4,2)$ is given by
the addition of the end point (4,2) to $P_e(3,1)$ 
not to $P_m(3,1)$.
This is because the global widths of both paths to $(4,2)$,
the path through $P_e(3,1)$ and the path through $P_m(3,1)$,
are the same as 3. 
Therefore, for a DPRMCF, we need information on the minimum 
height $y_{\min}$ and the maximum height $y_{\max}$ 
of the path as well as the potential energy of the path. 
In other words, the minimum energy $E_m(x,y)$ up to the site
$(x,y)$ is not enough but we need to know the minimum energies 
$E_m(x,y,y_{\min},y_{\max})$ of the paths to $(x,y)$ for all
different combinations of $y_{\min}$ and $y_{\max}$.
The minimum energy $E_m(x,y,y_{\min},y_{\max})$ can be 
calculated from the minimum energies 
$E_m(\xm,\ym,y'_{\min},y'_{\max})$
and $E_m(\xm,\yp,y'_{\min},y'_{\max})$
for a proper combinations of $y'_{\min}$ and $y'_{\max}$ values
and the ground state energy of Eq.~(\ref{eq.h}), $E_g(N)$ is 
obtained as the minimum of $E_m(N,y,y_{\min},y_{\max})$,
over $y$, $\ymi$, and $\yma$,
\beqa
 E_g(N) \a= \min_{y,y_{\min},y_{\max}} E_m(N,y,y_{\min},y_{\max}).
\label{eq.tr.a}
\eeqa
In principle, one can construct an algorithm to find
ground state energy and its path based on 
Eq.~(\ref{eq.tr.a}) but it generally requires 
$\cO(N^4)$ memory and time. Although its ``effective'' time 
complexity can be reduced $\cO(N^3)$ (see below), 
the $\cO(N^4)$ memory requirement puts a strong upper bound on
the sizes of the systems to be investigated. 

In this paper, we use a novel algorithm which uses only $\cO(N^2)$ 
memory but is more efficient than the algorithm using
Eq.~(\ref{eq.tr.a}). 
As before, we imagine the set $Y_{\ymi,\yma}$ 
of paths whose minimum and maximum heights 
are $\ymi$ and $\yma$ but we first consider the path with 
the minimum potential energy,
\beqa
 E^0_m(N,y_{\min},y_{\max})
	\a= \min_{\{y(x)\} \in Y_{\ymi,\yma}}
\left[\sum_{x=1}^N \exyx \right], 	\nnn
\eeqa
instead of the minimum total energy.
Since the minimum total energy $E_m(N,y_{\min},y_{\max})$ for
the given $\ymi$ and $\yma$ values, is simply given by 
\beqa
 E_m(N,\ymi,\yma)
	\a=  E^0_m(N,\ymi,\yma) \hspace{25mm} \nnn
	&& + \ \epsilon\, (y_{\max} - y_{\min} +1)^\alpha,
	\hspace{5mm}
\eeqa
the ground state energy $E_g(N)$ can be
obtained by 
\beqa
 E_g(N)
	\a= \min_{y_{\min},y_{\max}}
	E_m(N,y_{\min},y_{\max}) \hspace{20mm} \nnn
	\a= \min_{y_{\min},y_{\max}}
	\left[\, E^0_m(N,y_{\min},y_{\max}) \right. \nnn
	& & \hspace{15mm} 
		+ \ \left. \epsilon\, (y_{\max} - y_{\min} +1)^\alpha\,
	\right],
	\hspace{-20mm}
\label{eq.eg.a}
\eeqa
from the minimum potential energies. 
Note that we always have $\ymi \le 0$ and $\yma \ge 0$ 
since we set $y(0)=0$ as the anchored boundary. 
Therefore, in general, we need to calculate 
$E^0_m(N,\ymi,\yma)$ for $N^2$ different
combination of $\ymi\in\{0,-1,\ldots,-N\}$
and $\yma\in\{0,1,\ldots,N\}$ 
to get the ground state energy $E_g(N)$.
However, for a confining energy with positive $\epsilon$ and $\alpha$,  
we do not have to look for the paths with 
$\yma \ge \Wao$ or $\ymi \le -\Wao$ where
$\Wao$ is the width of the minimum potential path $P^0_g$,
which minimizes $E_{RM}$ of Eq.~(\ref{eq.erm}).
Let $\ymi^0$ and $\yma^0$ be the minimum and the 
maximum heights of 
the minimum  potential path $P^0_g$. 
Then $E^0_m(N,\ymi^0,\yma^0)$ is
the minimum of the first term in the square bracket
of Eq.~(\ref{eq.eg.a})
and $\Wao = \yma^0 - \ymi^0 + 1$. 
The total energy of a path, whose
$\yma \ge \Wao$ or $\ymi \le -\Wao$,  
cannot be smaller than that of $P^0_g$
since its confining energy is 
larger than $\epsilon \Wao^\alpha$
in addition to the fact that its
potential energy is lager than 
$E^0_m(N,\ymi^0,\yma^0)$. Therefore,
for a given $\eta(x,y)$, the ground state energy $E_g(N)$ of 
Eq.~(\ref{eq.eg.a}) can be obtained from 
the minimization of $E_m(N,\ymi,\yma)$ 
over $\ymi\in\{0,-1,\ldots,-\Wao+1\}$
and $\yma\in\{0,1,\ldots,\Wao-1\}$.

Now, let us introduce a simple way to calculate 
$E^0_m(N,\ymi,\yma)$ for the $(\ymi,\yma)$ pairs
needed for the minimization of Eq.~(\ref{eq.eg.a}). 
We use the conventional TMA for the DPRM problem
but with a series of modified site potentials,
\beqa
 \exyt(x,y) \a= 
	\left\{\begin{array}{ll}
	\exy &\hspace{5mm} \mbox{for\ }  \yL \le y \le \yU, \\
	\infty &\hspace{5mm} \mbox{otherwise,} 
	\end{array}
	\right. 
 \label{eq.exyt}
\eeqa
for $\yL > - \Wao$ and $\yU<\Wao$.
Note that we cannot obtained the minimum energies
$E^0_m(N,\ymi,\yma)$ for all pairs of $(\ymi,\yma)$
with $\ymi> -\Wao$ and $\yma<\Wao$
by simply applying TMA to the potential $\exyt$, 
since $(\ymi,\yma)$ of 
the minimum potential path $\pglu$ of $\exyt$
is not necessary equal to $(\yL, \yU)$. 
As an example, Fig.~{\ref{fig.tran.mat}(d) shows 
$\pglu$ for $(\yL, \yU)=(-2,1)$.
The minimum height of the $\pglu$, 
$y_{\min}=-2$ is equal to $\yL$ but the maximum height
of it, $y_{\max}=0$ is not equal to $\yU=1$.
Therefore, the minimum path energies for 
the two modified random potentials
$\tilde{\eta}_{-2,0}$ and $\tilde{\eta}_{-2,1}$
are the same as $E^0_m(N,-2,0)$ and we cannot obtain 
$E^0_m(N,-2,1)$ by the TMA with $\exyt$. 
However, this means that
$E^0_m(N,-2,1)$ is larger than $E^0_m(N,-2,0)$
and we can safely exclude $E_m(N,-2,1)$
from the candidates of the $E_g(N)$
since $E_m(N,-2,1)$ must be larger than
$E_m(N,-2,0)$.
In general, we cannot obtain $E_m(N,\yL,\yU)$
by the TMA with $\exyt$ 
if $(\ymi,\yma)$ of $\pglu$ is not equal 
to $(\yL,\yU)$ but we can exclude 
$E_m(N,\yL,\yU)$ in the minimization of 
Eq.~(\ref{eq.eg.a}) since this means
$E_m(N,\yL,\yU)$ is always larger than
$E_m(N,\ymi,\yma)$.
If, the $(\ymi,\yma)$ value is not equal to $(\yL,\yU)$,
$E^0_m(N,\yL,\yU)$ is larger than 
$E^0_m(N,\ymi,\yma)$ in addition to 
$(\yU-\yL+1)^\alpha>(\yma-\ymi+1)^\alpha$.
In other words, $E_g(N)$  
can be obtained by finding the minimum of 
$E_m(N,\ymi,\yma)$ over only the $(\ymi,\yma)$ 
pairs which can be obtained by the TMA with $\exyt$
for $\yL > - \Wao$ and $\yU<\Wao$.

Let us summary our algorithm to find the ground state
for a DPRMCF. For a given $\exy$, we first find the 
minimum potential path $P^0_g$ without the
confining energy term by using the conventional TMA
for a DPRM and calculate $\Wao$. 
This can be done in $\cO(N^2)$. 
Then we find $P^0_{\yL,\yU}$ by applying the TMA
to a series of the modified potentials $\exyt$ 
for $\yL > - \Wao$ and $\yU<\Wao$, 
and measure their potential energies $E^0_m(N,\ymi,\yma)$
and widths $\Wa=\yma-\ymi+1$.
For a given $\yL$ and $\yU$, $P^0_{\yL,\yU}$ can be
obtained in time of order $N(\yU-\yL)$.
Since $\Wao \sim N^{2/3}$,
all $\pglu$ can be obtained in $\cO(N^3)$
on average.
The ground state energy $E_g(N)$ is then given by the minimization of
$E_m(N,\ymi,\yma)$ through Eq.~(\ref{eq.eg.a}) and the ground state
path is given by the corresponding minimum energy path.
Note that the ground state energies and their
paths for all positive $\epsilon$ and $\alpha$ 
values can be obtained using Eq.~(\ref{eq.eg.a}) 
once we get $\pglu$ for  $\yL > - \Wao$ and $\yU<\Wao$. 

\section{Numerical Results}

\begin{figure}[t!] 
\includegraphics[width=7cm]{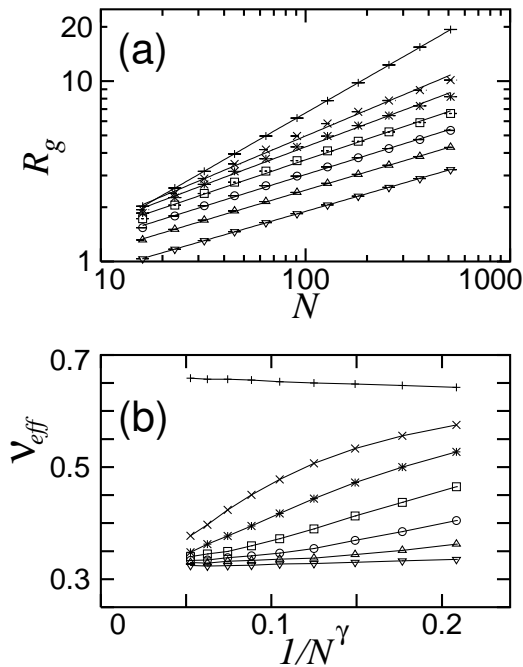} 
\caption[0]{
(a) Radius of gyration $R_g$ for 
$N$=16, 23, 32, 45, 64, 92, 128, 256, 362, and 512 systems 
with a harmonic confining energy $E_c=\epsilon W^2$ ($\alpha = 2$). 
From the top, $R_g$ for 
$\epsilon$=0, 0.001, 0.002, 0.004, 0.008, 0.016 and 
0.04 cases are shown. Fitting lines are 
in the form of $R_g = a N^{\nu}$
with $\nu$=0.65, 0.48, 0.43, 0.38, 0.36, 0.34 
and 0.33 respectively from the top.  
(b) Effective roughness exponents $\ne$
for $\epsilon$=0, 0.001, 0.002, 0.004, 0.008, 0.016 and 
0.04 values are plotted against $1/N^{\gamma}$
with $\gamma=0.5$ from the top. The uppermost curve of $\epsilon=0$
goes to the known value of $\nu_o=2/3$ but all the other curves go 
to much smaller value around 1/3 as $N$ goes to infinity.
}
\label{fig.rg.nu}
\end{figure}

We first consider a harmonic confining energy of
$\epsilon \Wa^2$, that is, the $\alpha = 2$ case
with the random potentials $\eta(x,y)\in[0,1]$
taken from the uniform distribution. 
We simulate a random potentials $\eta$ using the 
computer-generated pseudo random numbers~\cite{ran-gen} 
and find the ground state path 
of Eq.~(\ref{eq.h}) using the algorithm presented 
in the previous section.
Then, we measure the square 
width $R_\eta^2$ (the square bracket of the Eq.~(\ref{eq.rg}))
and the first $E_\eta$ and the second 
$E_\eta^2$ moments of the energy for the ground state path 
for the given $\eta$. 
The radius of gyration $R_g$ and $\Delta E$ 
are then obtained as $R_g = \sqrt{\ave{R_\eta^2}}$
and $\Delta E = \sqrt{\ave{E^2}-\ave{E}^2}$
where $\ave{A}$ mean the average over 
different realization of random potentials.
We use eight million different realizations ($\Omega = 8 \times 10^6$)
of random potentials for $N$=16, 23, 32, 45, 64, 92 and 128 systems 
and four, two and one million different realizations 
for $N$=256, 362 and 512 systems respectively
to obtain the average values. 
The average over this large number of different random potentials
makes the statistical error bars smaller than the sizes of the symbols
most cases except the effective energy fluctuation 
exponents $\oe$ shown in 
Fig.~\ref{fig.de.ome}(b) and 
Fig.~\ref{fig.rg.nu.a1}(d) later. 

Figure~\ref{fig.rg.nu}(a) shows 
the radius of gyration $R_g$ for the harmonic confining energy 
$E_c = \epsilon W^2$ as a function of polymer length $N$
for $\epsilon$=0, 0.001, 0.002, 0.004, 0.008, 0.016, and 0.04 cases.
For each $\epsilon$ value, $R_g$ lies on a straight line in a 
log$-$log scale plot indicating $R_g(N) \sim N^{\nu}$.
The least $\chi^2$ fits of $R_g(N) \sim N^{\nu}$ give
$\nu$=0.66, 0.48, 0.43, 0.38, 0.36, 0.34 
and 0.33 
for $\epsilon$=0, 0.001, 0.002, 0.004, 0.008, 0.016, and 0.04 cases
respectively. The ``roughness'' exponents for $\epsilon>0.04$ are
almost identical to those of the $\epsilon=0.04$ case 
unless $\epsilon$ is very large where 
the finite size effect is strong.

To estimate $\nu$ values for $N\rightarrow\infty$, 
we calculate $N$ dependent effective roughness exponent $\ne$ defined
by the successive slopes in the log$-$log plot. We use
neighboring three points to get the local slope, {\it i.e.},
$\ne(N_k)$ is obtained as the slope of the least $\chi^2$ fit 
using
the three data $R_g(N_{k-1})$, $R_g(N_{k})$, and $R_g(N_{k+1})$,
where $N_k$ are the system sizes in an ascending order,
$16=N_1<N_2<\cdots<N_{10}=512$.
The roughness exponent $\nu$ is estimated by extrapolating
$\ne$ in the infinite size limit with 
a plot $\ne$ against $1/N^\gamma$.
In Fig.~\ref{fig.rg.nu}(b),
we choose $\gamma=0.5$ which characterizes the trend of $\ne$ for 
large $N$ well.  
From the figure, we see that our numerical measurement 
of $\nu$ for $\epsilon=0$ is consistent with the 
known value $\nu_o=2/3$ of the DPRM. 
However, it is clear that the roughness exponents 
of the DPRMCF ($\epsilon>0$) is not equal to $\nu_o$.
It is difficult to extract the definite $\nu$ values
from our numerical data but we speculate that $\ne$
goes to the conjectured value of
1/3 for all $\epsilon>0$  
as the system size $N$ goes to infinity.

\begin{figure}[t!] 
\includegraphics[width=7cm]{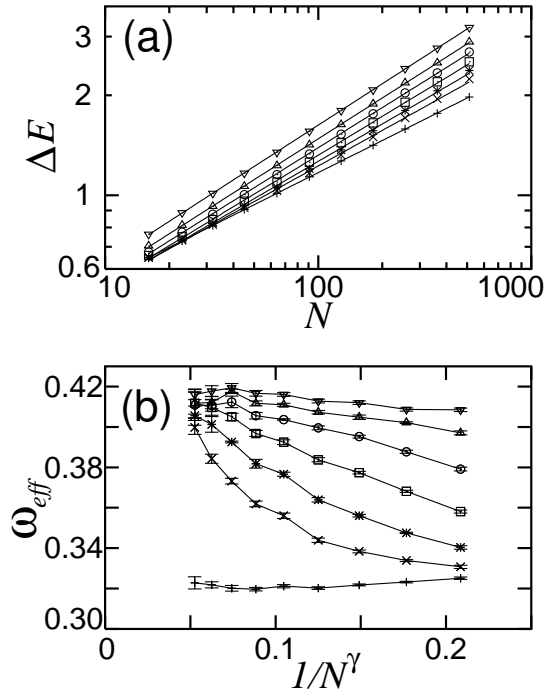} 
\caption[0]{
(a) Measured energy fluctuation
$\Delta E$ for the systems of sizes 
$N$=16, 23, 32, 45, 64, 92, 128, 256, 362, and 512
for $\alpha = 2$. 
From the bottom, $\Delta E$ for 
$\epsilon$=0, 0.001, 0.002, 0.004, 0.008, 0.016 and 
0.04 cases are shown. Fitting lines are 
in the form of $\Delta E = a N^\omega$
with $\omega$=0.32, 0.36, 0.37, 0.39, 0.40, 0.41 
and 0.41
respectively from the bottom.  
(b) Effective exponents $\oe$
for the above 7 different values of 
$\epsilon$ are plotted against $1/N^{\gamma}$ 
with $\gamma=0.5$. As $N$ goes to infinity,
the lowest curve of $\epsilon=0$
goes to the known value of $\omega_o=1/3$ but all the other curves go 
to much larger value around the predicted $\omega$ of $5/12$.
}
\label{fig.de.ome}
\end{figure}

Figure~\ref{fig.de.ome}(a) shows the energy fluctuation 
$\Delta E$ as a function of polymer length $N$
for the above seven $\epsilon$ values. 
As in the case of $R_g(N)$, 
for each $\epsilon$ value, $\Delta E$ lies on a straight line 
in a log$-$log scale plot indicating a power law increase.
The least $\chi^2$ fits of $\Delta E(N) \sim N^\omega$ give
$\omega$=0.32, 0.36, 0.37, 0.39, 0.40, 0.41 
and 0.41
for $\epsilon$=0, 0.001, 0.002, 0.004, 0.008, 0.016, and 0.04 cases
respectively. We estimate the energy fluctuation exponent
$\omega$ as the $N\rightarrow\infty$ limit of
the effective exponents $\oe$ as for the roughness exponent.
The effective exponent $\oe$ is defined as 
the successive slopes in the log$-$log plot of 
Fig.~\ref{fig.de.ome}(a)
and obtained by the least $\chi^2$ fit with
the three neighboring points. 
In Fig.~\ref{fig.de.ome}(b), the effective exponents
are plotted against $1/N^\gamma$ with $\gamma=0.5$ as before. 
From the figure, we see that our numerical measurement 
of $\omega$ for $\epsilon=0$ is consistent with the 
known value $\omega_o=1/3$ of the DPRM.
For $\epsilon\not=0$, all curves seem to go 
the conjectured value of $\omega=\frac{5}{12}\approx 0.417$.

\begin{figure}[t!] 
\includegraphics[width=7cm]{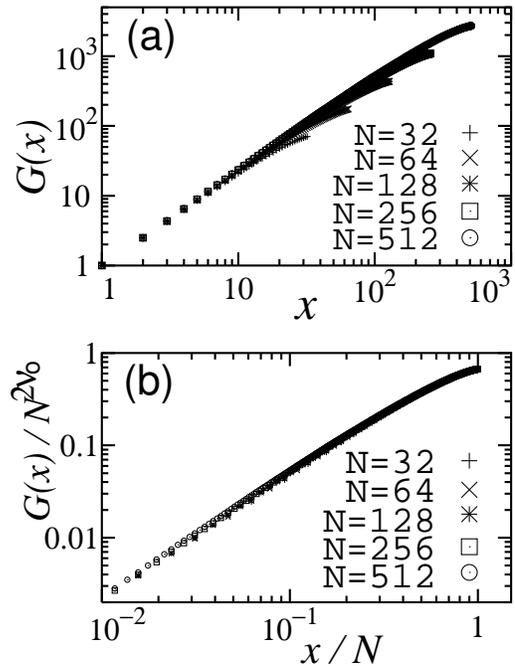} 
\caption[0]{
(a) $G(x)$ for the systems of sizes 
$N$=32, 64, 128, 256, and 512 with no 
confining force $\epsilon=0$.
(b) A rescaled height correlation $G(x)/N^{2\nu_0}$
is plotted 
against rescaled distance $x/N$ with $\nu_0=2/3$
in log scale. 
}
\label{fig.gr.a}
\end{figure}

In addition to the radius of gyration, we also measure the 
height (transverse position) correlation function $G(x;N)$ to 
check the ``window'' argument.
The correlation function $G(x;N)$
is defined by the mean square height of the site $x$
of a polymer of length $N$,
\beqa
 G(x;N)	\a= \AVE{y(x)^2}	\nnn
	\a= \frac{1}{\Omega} \sum_{\{\eta(x,y)\}} 
		y^2_\eta(x)	
\eeqa
where $\Omega$ and $y_\eta(x)$ are the same as in the 
Eq.~(\ref{eq.rg}). 

\begin{figure}[t!] 
\includegraphics[width=7cm]{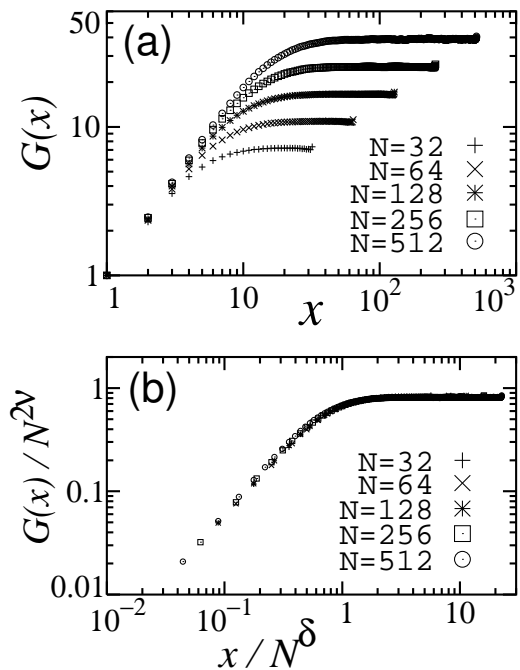} 
\caption[0]{
(a) The correlation function $G(x)$ for the systems 
of sizes $N$=32, 64, 128, 256, and 512 with a 
confining energy $E_c= 0.04 W^2$.
(b) The rescaled height correlation $G(x)/N^{2\nu}$
is plotted against rescaled distance $x/N^\delta$ 
in a log scale with $\nu=1/3$ and $\delta=1/2$.
}
\label{fig.gr.b}
\end{figure}

Figure~\ref{fig.gr.a}(a) shows the $G(x;N)$ 
for the DPRM ($\epsilon=0$) for 
$N=32$, 64, 128, 256 and 512 systems.
All the data
collapse to a single curve with $G(x)\sim x^{2/3}$ except the points
near or at the boundaries at $x=N$. Note that the ground state path 
$y_\eta(x)$ for a given random potential $\eta$ cannot be determined
locally even in the case of $\epsilon=0$. In other words, the minimum
energy path up to $x$ is different from the sub-path up to $x$ of the
minimum energy path of length $N>x$. 
Due to the global nature of the ground state path determination, 
$G(x;N)$ deviate form the infinite size behavior for 
a finite portion of $N$ even for the free 
boundary condition~\cite{loc.bc}. 
Yet, the height correlation function follows the scaling relation,
\beqa
 G(x;N)	\a= N^{2\nu_o} g_o(x/N)
\eeqa
where $g_o(u)$ increases as $g_o(u)\sim u^{2\nu_o}$ unless 
$u$ is very close to 1 where the boundary effect exists. 
When we rescale the correlation function $G(x;N)$ by $N^{2\nu_o}$
with $\nu_o=2/3$ and $x$ by $N$,
all the data collapse to a 
single curve as shown in Fig.~\ref{fig.gr.a}(b). 

However, the correlation function $G(x;N)$ for 
$\epsilon>0$ shows qualitatively different 
behaviors from the DPRM case. 
When there is an energy term associated with the 
global width, there seems to be another length scale over 
which the correlation function saturated.
Figure~\ref{fig.gr.b}(a) shows the $G(x;N)$ for $\epsilon=0.04$
for $N=32$, 64, 128, 256 and 512 systems.
As $x$ increases, the correlation functions increase algebraically
only for $x \le l(N)$ and then remains as constant values. 
We rescale the correlation functions $G(x;N)$ by $N^{2\nu}$
and $x$ by $N^\delta$ with the conjectured values of 
Eq.~(\ref{eq.nu}), $\nu=1/3$ and $\delta=1/2$ 
and plot $G(x;N)/N^{2\nu}$ against $x/N^\delta$
in Fig.~\ref{fig.gr.b}(b). All data collapse to a single 
curve, implying a new scaling law 
\beqa
 G(x;N)	\a= N^{2\nu} g(x/N^\delta).
\eeqa
The scaling function $g(u)$ increases as $g(u)\sim u^{2\nu_o}$
for $u < 1$ and then becomes a constant for 
$u > 1$. In other words, $G(x;N)$ increase
as $x^{2\nu_0}$ for $x<N^\delta$ and
reaches a constant value for $x>N^\delta$.
Note that the scaling function grows 
algebraically with exponent $2\nu_0$ 
not with $2\nu$ supporting the assumption
that our polymer behaviors like a DPRM 
up to the window size $N^\delta$ and then
feels the global constraints of the 
confining force over the window size. 

\begin{figure}[t!] 
\TwoFigPlace{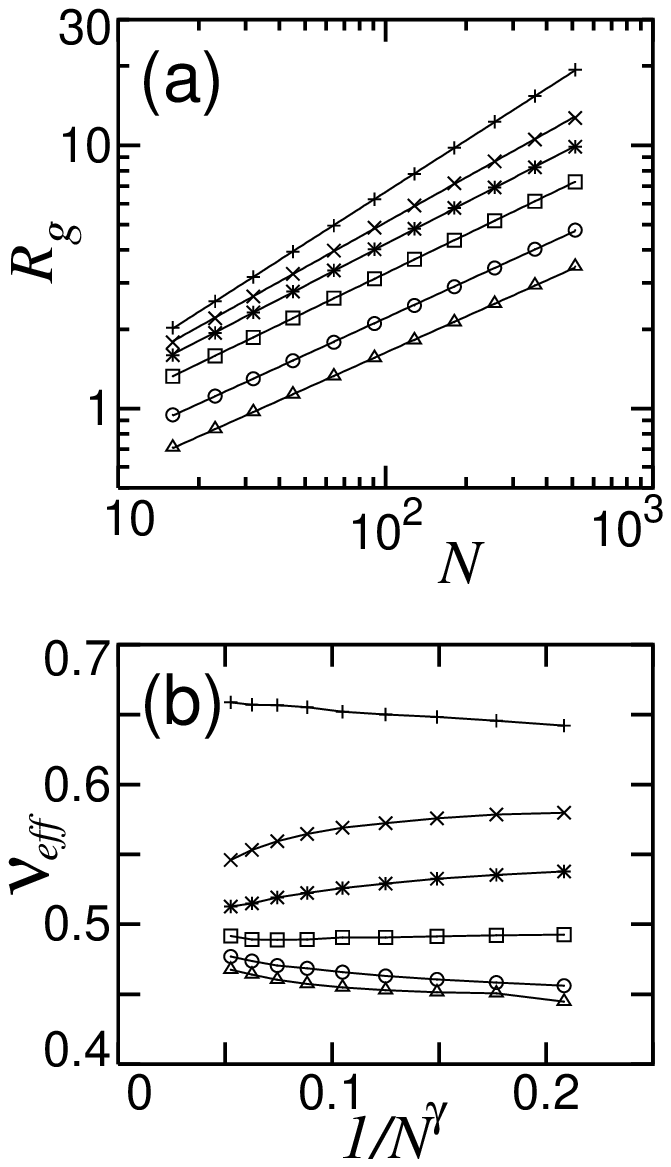}{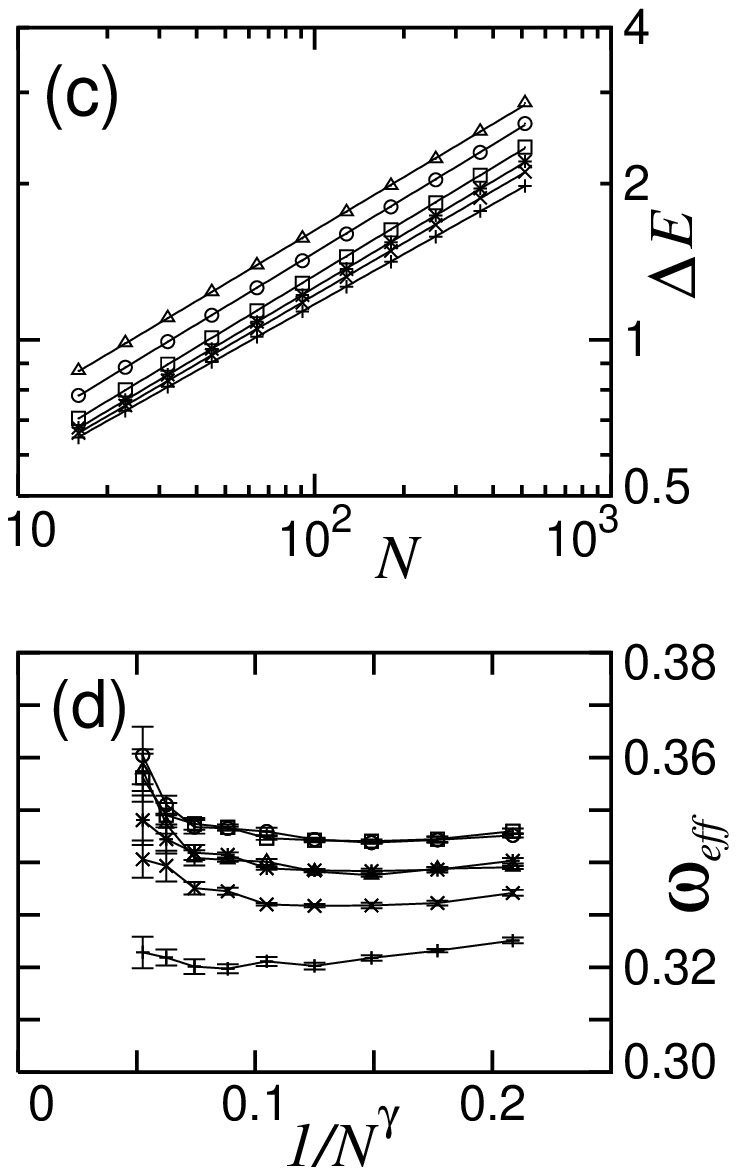}{3.2}{60}{60}{-53}{65}{-60}{-60}
\caption[0]{
(a) Measured $R_g$ for the systems of sizes 
$N$=16, 23, 32, 45, 64, 92, 128, 256, 362, and 512
for $\alpha = 1$. 
From the top, $R_g$ for 
$\epsilon$=0, 0.05, 0.1, 0.2, 0.5, and 1.0 cases are shown. 
Fitting lines are in the form of $R_g = a N^{\nu}$
with $\nu$=0.65, 0.57, 0.53, 0.49, 0.46 and 0.45 from the top.  
(b) Effective roughness exponents $\ne$ for the above 7 
different values of 
$\epsilon$ are plotted against $1/N^\gamma$ with $\gamma=0.5$.
As the system size goes to infinity, the effective exponents
seems to go to the conjectured value $0.5$ for all $\epsilon\not=0$.
(c) Energy fluctuations $\Delta E$ are plotted against $N$.
Fitting lines are in the form of $\Delta E = a N^{\omega}$
with $\omega$=0.32, 0.33, 0.34, 0.35, 0.35 and 0.35 
in an ascending order or epsilon values 
from the bottom.  
(d) Effective energy fluctuation exponents $\oe$ 
against $1/N^{\gamma}$ with $\gamma=0.5$. The lowest curve 
is for the $\epsilon=0$ case. 
}
\label{fig.rg.nu.a1}
\end{figure} 

We perform a series of simulations for other values of $\alpha$ 
and obtain the numerically results consistent 
with the conjecture of Eq.~(\ref{eq.nu}).  
Figure~\ref{fig.rg.nu.a1} shows the radius of gyration $R_g$ 
and the energy fluctuation $\Delta E$ and their 
effective exponents $\ne$ and $\oe$ for the 
confining energy of $W_c = \epsilon W$ ($\alpha=1$).
The radius of gyration $R_g$ lies on a straight line in a 
log$-$log scale plot as before and 
the least $\chi^2$ fits of $R_g(N)\sim N^{\nu}$ give
$\nu$=0.65, 0.57, 0.53, 0.49, 0.46 and 0.45 
for $\epsilon$=0, 0.05, 0.1, 0.2, 0.5, and 1.0 respectively. 
Note that the $\nu$ values from the least $\chi^2$ fits 
(using $R_g$ for $N\le 512$) are smaller than the conjectured
value of 1/2 for some large epsilon values while 
those for small epsilon values are larger than the conjectured
values unlike the case of $\alpha=2$. Yet, as the system size
goes to infinity, all the effective roughness exponents $\ne$ seem
to go to the conjectured value of 1/2 for $\epsilon>0$ while
$\ne$ for $\epsilon=0$ goes to the known value of 2/3
(see Fig.~\ref{fig.rg.nu.a1}(b)). The energy fluctuation $\Delta E$ 
is also measured and the effective energy fluctuation exponents $\oe$
are calculated from them. As shown in Fig.~\ref{fig.rg.nu.a1}(c), 
the energy fluctuation $\Delta E$ increases as 
$\Delta E \sim N^{\omega}$
with $\omega$=0.32, 0.33, 0.34, 0.35, 0.35 and 0.35 
for $\epsilon$=0, 0.05, 0.1, 0.2, 0.5, and 1.0 respectively. 
From these data, it seems to be difficult to distinguish 
the energy fluctuation exponent of DPRMCF from that of
DPRM. The analysis of the effective exponents $\ne(N)$ 
provides somewhat better diagnosis. In Fig.~\ref{fig.rg.nu.a1}(d)
we plot $\ne$ against $1/N^{0.5}$ as before.
The error bars of $\oe$ for $\alpha=1$ are relatively large
as shown in the figure although we obtain the effective 
exponent with more than million different random potentials.
The statistical error of one millionth
in measuring the second moment average $\ave{E^2}$
gives rise to the error bars of the symbol size
in Fig.~\ref{fig.rg.nu.a1}(d). Due to these limitations, 
we cannot extract the definite value of $\omega$ for 
$\epsilon>0$ from our simulations 
but the numerical results does not seem to exclude
the conjectured value of $\omega=3/8=0.375$.
Our simulations on other $\alpha$ values such as 
$\alpha=1.5$ and $\alpha=3$ also give the consistent
results with the conjecture of Eqs.~(\ref{eq.nu}) 
and~(\ref{eq.omega}). 

\section{Concluding remarks}
We consider the scaling behavior of a directed polymer
in a 2D random media with confining energy 
$E_c = \epsilon W^\alpha$
and find that the roughness exponent $\nu$ and the energy fluctuation 
exponent $\omega$ are given by $\nu = 1/(\alpha+1)$ and 
$\omega = (2\alpha+1)/4(\alpha+1)$ respectively. 
These results can be 
understood by assuming that a polymer of length $N$ 
behaves like a DPRM up to the window size $l(N)\sim N^\delta$ 
and then feels the confining energy over the window size. 

We have only considered the scaling behavior of the zero 
temperature ground state polymers. We know the 
finite temperature behaviors of the polymers only for some limiting 
cases. For DPRM where confining energy term is absent,
zero temperature pinned phase is the fixed point so that 
polymers at any finite temperature shows the same exponents with those
at the zero temperature. On the other hands, if there is
only confining energy (without random
potential) there are three phases as temperature changes. 
At zero temperature, the polymer becomes a
straight line with width 1 and therefore $\nu=0$ while it
becomes a random walk with $\nu=1/2$ at infinite temperature.
At nonzero finite temperature, especially for $\alpha=1$ where
the confining energy is given by $E_c = \epsilon W$, the 
polymer configurations are identical to the self flattening
surface~\cite{Noh01PRE} whose roughness exponent $\nu$ is $1/3$. 
When there are both random potential and confining energy
terms, we do not know the scaling behaviors of the polymers
at finite temperature where our algorithm cannot be applied.
Further investigations are needed to explore full phase diagram
of DPRMCF over general temperature and dimensions. 

\begin{acknowledgments}
We thank J. M. Kim, J. D. Noh and H. Choi for useful comments
and KIAS for the hospitality during the visit. 
This work was supported by the Korea Research Foundation
Grant (KRF-2004-015-C00188) and the numerical calculations in this
work were performed by using the supercomputing resources of Korea
Institute of Science and Technology Information.   
\end{acknowledgments}
 

\begin{thebibliography}{10}
\bibitem{ran-gen}
To check the possibility of the systematic errors from the computer generated
  pseudo random numbers, we use four different random number generators,
  drand48 in the standard library and ran1, ran2 and ran3 in the numerical
  recipes for the simulations for the systems with $N\le 128$. With each random
  number generator, we perform two million different simulations and confirm
  that all the measured quantities including $R_g$, $\nu_{\!e\!f\!f}$, $\Delta
  E$, and $\omega_{e\!f\!f}$ coincide within the statistical errors. For
  $N>128$, we use drand48.
\bibitem{loc.bc}
When the path configurations are locally determined as in a restricted solid on
  solid model (RSOS), the boundary effects on the height correlation function
  $G(x;N)$ exist only for the sites in a finite range from the boundary ($N-x <
  C$) unless the boundary condition puts a global constraint. For example,
  $G(x;N)$ of a RSOS model with the free boundary condition has no finite size
  effect~\cite{Jeong03PRE} unlike that with the periodic boundary condition
  where the global sum of the local slops must be zero.
\end{thebibliography}

\end{document}